# Taking a step back and looking at the "superconducting dome" from a distance

Andreas Schilling

Department of Physics, University of Zurich, Winterthurerstrasse 190 CH-8057 Zurich Switzerland

**Abstract**

In my short essay in honor of Karl-Alex Müller, I would like to deal with some aspects that have been on my mind since the beginning of my own research. Some of the facts and questions mentioned here have been known or clarified for a long time, others not. For example, I think that the fact that superconductivity occurs in many cuprates almost spontaneously close to the optimum doping level has been insufficiently (or not at all) investigated. Much research has been done in the regions outside the optimal doping, be it underdoped or overdoped. I would like to deviate from the "mainstream" for a moment and ask whether these regions are really so relevant for the occurrence of superconductivity, or whether they simply serve to interfere with it - a thought which I am sure has been expressed by others and which would certainly lead (or had already led) to lively discussions with Karl-Alex Müller.

## 1. Introduction

While it is undisputed that scientific progress is ultimately based on precise scientific data, the interpretation of such data can be influenced by factors other than purely inductive reasoning and can therefore sometimes lead to a dead end. Prior to the discovery of high-temperature superconductors by Karl-Alex Müller and Georg Bednorz [1], research in the field of superconductors had, in a sense, reached such a dead end. The critical temperatures $T_c$ of metallic superconductors had reached a maximum around 23 K in the A15 family, and the underlying mechanism of superconductivity in these materials was essentially known and well-accepted, so no further dramatic improvements could be expected in the future. However, there already existed systems that deviated from the "standard model" for superconductivity, such as heavy-electron compounds [2] and some oxides that did not really fit the general picture [3-5]. Karl-Alex Müller and Georg Bednorz dared to take a step back and took the challenge to investigate the hypothesis that superconductivity in oxides is not limited to only a few examples such as doped $SrTiO_3$ [3], $LiTiO_4$ [4] or $BaBiO_3$ [5]. Realizing that superconductivity in these oxides can only be achieved by mixed-valent metal atoms (induced by doping with additional charge carriers into the stoichiometric "parent" compounds), they finally discovered superconductivity in mixed-valent cuprates with a perovskite-like structure, although they initially took an unsuccessful detour via related nickel oxides [6], which only recently attracted much attention from the scientific community again with the hoped-for discovery of superconductivity in them [7-10].

At present, the study of these cuprates also threatens to reach an impasse, since the critical temperatures have stagnated for decades around 133 -138 K at ambient pressures in mercury-based copper oxides [11,12], and a "smoking-gun proof" by a theory explaining their extraordinarily high transition temperatures is still lacking. Instead of repeating facts and arguments from my own published papers, in my short essay in honor of Karl-Alex Müller I will also try to take a step back and pick out peculiarities that I noticed early on in our experiments with cuprates, for which I have not yet found a satisfactory explanation, and which may yet stimulate one or two fruitful short discussions.

## 2. Why is it so easy to reach high critical temperatures in some cuprate superconductors?

I chose this provocative title because it reflects reality. To pique the reader's curiosity, I can say that our discovery of superconductivity above 130 K in a mercury cuprate many years ago [11] would not have been possible if it had not behaved exactly as announced in the title above. At my current and at Karl-Alex Mueller's former home institute, it is common that every year, that some high-school students do a short research project on superconductivity. The first task for most of them is to prepare high-temperature superconducting samples for later measurements or for the design of another experiment, since they are easy to synthesize if one knows the exact composition in advance. A similar task would be very difficult for intermetallic compounds, and perhaps somewhat unsatisfactory for elemental superconductors because of their modest critical temperatures at ambient conditions (up to $T_c \approx 9.3$ K in niobium) compared to high-temperature superconductors. Some readers may be aware that a simple method for preparing copper-oxide compounds is to intimately mix appropriate amounts of corresponding metal oxides or nitrates, grind them repeatedly, and bake them (often just in air) until a sample containing the desired phase is obtained. Because of the relatively short reaction times, we usually recommend that students prepare $YBa_2Cu_3O_{7-\delta}$ with a maximum critical temperature of $T_c \approx 92$ K. The role of the final annealing step is to ensure that the oxygen content reaches a level such that the copper acquires a mixed average valence of $\approx +2.15$ to $+2.18$, resulting in critical temperatures near the maximum of the respective "superconducting dome" (see Fig. 1a) [13].

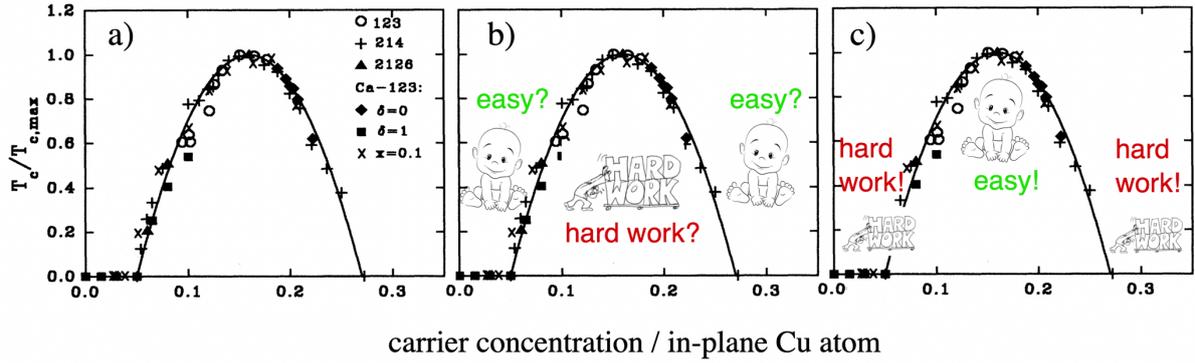

*Fig. 1.* (a): The "superconducting dome", normalized to the maximum critical temperature $T_{c,max}$ for several cuprate superconductors (figure reproduced with permission from Ref. [13]). (b) and (c): Scenarios illustrating the difficulty to achieve the maximum critical temperature.

If you have never performed such an experiment (or any experiment at all), you might think that reaching a reasonably high critical temperature is very tricky and requires precise fine-tuning of all reaction parameters, such as annealing temperature or oxygen partial pressure, and that reaching the maximum critical temperature is even more difficult (Fig. 1b). However, the opposite is true (Fig. 1c). In Fig. 2, I show representative plots for the prototype materials $YBa_2Cu_3O_{7-\delta}$ [14] and $Bi_2Sr_2CaCu_2O_{8+\delta}$ [15], where the critical temperatures or oxygen contents are given as functions of the respective reaction temperatures and oxygen partial pressures. It is clear that there must exist qualitatively similar diagrams not only for the closely related compound $Bi_2Sr_2CuO_{6+\delta}$, but also for thallium-based (Tl) and mercury-based (Hg) copper oxides such as $Tl_2Ba_2Ca_{n-1}Cu_nO_{2n+4+\delta}$ and $HgBa_2Ca_{n-1}Cu_nO_{2n+2+\delta}$ (with $n$ an integer number indicating the number of Cu-O sheets per formula unit), respectively, perhaps with the exception of $YBa_2Cu_4O_8$ with a critical temperature of $T_c \approx 80$ K, where the oxygen content is virtually invariable.

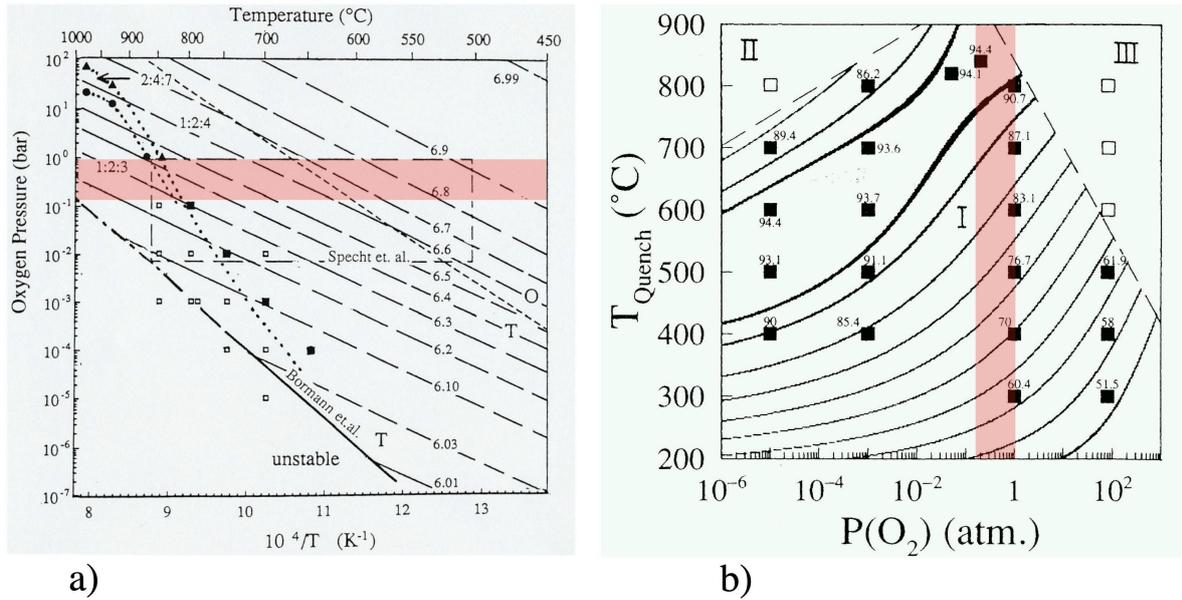

a) b)

*Fig. 2:* (a) Synthesis conditions (annealing temperatures and oxygen partial pressures) to obtain a certain oxygen content in $YBa_2Cu_3O_{7-\delta}$ (picture taken with permission from Ref. [14]). (b) Quenching temperatures and oxygen partial pressures to obtain a certain critical temperature for $Bi_2Sr_2CaCu_2O_{8+\delta}$ (graph taken with permission from Ref. [15]). The red shaded areas indicate standard laboratory atmospheres (oxygen partial pressures between ≈ 20% and 100%.)

The red shaded areas indicate oxygen partial pressures between air (≈ 20%) and ≈ 100% (which can be approximated by applying a flow of pure oxygen gas). When cooled from the typical reaction temperatures around 800-950 °C down to room temperature, the materials seem to automatically adopt an oxygen content that leads to a critical temperature rather close to the respective optimum value. It is the "$\delta$" that makes the difference here. Stoichiometric $YBa_2Cu_3O_6$ and $Bi_2Sr_2CaCu_2O_8$ would be antiferromagnetic insulators just like the prototype insulating perovskite $La_2CuO_4$, that became so famous only after the substitution of La by Ba by Karl-Alex Müller and Georg Bednorz led to their discovery. Moreover, both the thallium-based copper-oxide family and the mercury-based compounds with their record critical temperatures behave essentially the same: once you have synthesized the correct cationic ionic lattice structure at high temperatures, the oxygen content adjusts itself almost automatically upon cooling to room temperature to a non-stoichiometric value that guarantees a favourable mixed valency of copper to reach a high critical temperature near its optimum value. Can superconductivity be missed in these selected systems? Hardly. Nature seems to provide an almost free and effortless way to achieve a high $T_c$ in these materials. On the contrary, neither the hypothetical antiferromagnetic insulators with exact compositions $Tl_2Ba_2Ca_{n-1}Cu_nO_{2n+4}$ nor $HgBa_2Ca_{n-1}Cu_nO_{2n+2}$, have, to my knowledge, ever been successfully synthesized because it is either extremely difficult or even impossible to obtain them (Fig. 1c).

So how does nature help us achieve superconductivity near optimum doping for next to nothing? An important reason may be the presence of large voids within the heavy-element layers in the Bi, Tl and Hg based compounds that can host extra oxygen. In $YBa_2Cu_3O_{7-\delta}$, oxygen can easily be added to or removed from the so-called chain site, and stoichiometric $YBa_2Cu_3O_7$, although not at its optimum, happens to acquire a copper valency that still guarantees a $T_c$ well above liquid nitrogen temperature. Moreover, the relatively small grain size typically found in polycrystalline samples (of the order of tens of microns) allows the average oxygen content to equilibrate relatively quickly by diffusion (within minutes or hours), while it can take several weeks or even months in large single crystals [16]. Nevertheless, it seems almost miraculous that a certain oxygen level which leads to a very high $T_c$ value, and no other, often sets in as if by itself, which may or may not be purely coincidental.

To further illustrate, I may continue the anecdote about our work on the mercury-based cuprate $HgBa_2Ca_2Cu_3O_{8+\delta}$, and I admit that the first samples we had synthesised in 1993 [11] were a kind of crystallographic nightmare. Not only did they contain several superconducting phases, but also several impurities that made it impossible to identify the phase with the highest $T_c$ by X-ray analysis alone. This was finally achieved only through careful systematic scanning electron microscopy. If these synthesis experiments, in which we did not pay any attention at all to a special reaction atmosphere, had not immediately resulted in samples showing a strong Meissner effect already at 117 K, they might have simply ended up in the waste. Apparently, superconductivity had developed spontaneously, and, inspired by charts as shown in Fig. 2, we were subsequently able to catapult it to an even higher transition temperature around 133 K by annealing at moderate temperature in pure oxygen, which obviously helped to increase the copper valency somewhat by fine-tuning "$\delta$". We had tried this annealing procedure without any further information, as a shot in the dark, so to speak. It is interesting that precisely this $T_c$ value was reached and not another, which is very close to the maximum possible critical temperature in this system at ambient pressure. In other words, the material had almost effortlessly reached an optimal doping level without us knowing the exact annealing conditions for it.

Not to belittle Karl-Alex Müller's and Georg Bednorz' achievement, it must be emphasized at this point that $La_2CuO_4$ does not quite belong in the same category as the cuprates mentioned above. In the La-Cu-O system, the generation of a mixed valence is primarily achieved by a deliberately induced partial cationic substitution (e.g., by Ba or Sr instead of La), which must first be very carefully thought out, as Karl-Alex Müller and Georg Bednorz successfully did [1]. Nevertheless, it should be said that even unsubstituted $La_2CuO_4$ itself can become superconducting under suitable mild annealing conditions by intercalating oxygen into interstitial sites, resulting in samples containing a certain volume fraction of nearly optimally doped $La_2CuO_{4+\delta}$ with a $T_c$ up to $\approx$ 40 K, even when synthesized at 1 atm oxygen pressure [17,18].

In the above examples, it is the presence of extra oxygen that optimally dopes "underdoped" samples, and this may be due to some natural tendency to fill interstitial spaces with oxygen. However, there is also a counterexample. Li et al. have recently published a study on overdoped single-crystals of $La_{2-x}Sr_xCuO_4$ [19]. While the bulk critical temperature followed the trend of the "superconducting dome" on the overdoped side of the dome as shown in Fig. 1, a certain volume fraction with near-optimal $T_c \approx$ 38 K persisted up $x \approx$ 0.3 and finally vanished together with the complete disappearance of bulk superconductivity with $T_c \rightarrow$ 0 at even higher doping. An interpretation of a separation into components with optimal doping and overdoped components with lower $T_c$ is quite obvious here. If this is the case, the component with "optimum" critical temperature must have had an oxygen deficiency of the order of $\delta \approx$ – 0.06 to reach the desired copper valency around +2.18, thus indicating vacancies on the oxygen sites. The question remains as to why such separation occurs in the first place and why one of the components had a robust critical temperature just in close proximity to the optimal doping. It almost seems as if, in general, a copper valence near +2.15 to +2.18, usually set during synthesis or subsequent annealing of the material well above the critical temperature, is chemically preferable.

## 3. More on the "superconducting dome"

**a) The good things and the bad things for superconductivity**

To move on to another topic, I again quote Ref. [19] with the first sentence in the abstract: *"The interpretation of how superconductivity disappears in cuprates at large hole doping has been controversial"*. The way of plotting the "superconducting dome" as in Fig. 1 and interpreting it is very suggestive. It may imply that, using the western way of reading a text from left to right, superconductivity "emerges" on the left side of the dome and "disappears" on its right side. In other words, the psychology of reading the plot can lead to the conclusion that the "good things" for superconductivity are on the left and the "bad things" on the right side of the dome.

I assume that the first versions of the "superconducting dome" for high-$T_c$ superconductors were inspired by similar "domes" observed in previously known systems, such as heavy-electron compounds which are also intimately related to antiferromagnetism [20]. In these systems and if superconducting, $T_c$ can be varied and even suppressed upon changing a control parameter such as pressure, doping or magnetic field [21]. This is reminiscent of the situation in cuprates, whose parent compounds are antiferromagnets. Only when the hole carrier concentration is increased do they become superconducting, while the antiferromagnetic order is completely suppressed, although the "superconducting dome" is not directly adjacent to the antiferromagnetic region. Similar domes have also been seen in many other systems, such as in intercalated transition-metal dichalcogenides where $T_c$ varies as a function intercalant concentration, also often showing a maximum, and where superconductivity is usually regarded as being in competition with other types of orders such as the formation of charge-density waves [22]. Interestingly, hole-doped cuprates do also show charge order at low doping (i.e., on the left side of the dome). This is usually considered detrimental to the occurrence of superconductivity [23-27], as illustrated by the notoriously low Meissner fraction indicating incomplete spontaneous expulsion of magnetic flux. Keeping this in mind, may we speculate for a moment whether the "bad things" for superconductivity are rather on the left and the "good things" on the right side of the dome? It is remarkable that complications such as charge ordering or other types of peculiarities, such as the presence of the so-called pseudo-gap [28], are virtually absent at high enough hole-doping levels.

**b) "Overdoped" side**

To illustrate my point, let us suppose we were living in a world where a copper valence of +3 (i.e., a Cu $3d^8$ configuration) is very common, but a $Cu^{+2}$ ($3d^9$) is very difficult to achieve. This situation perhaps reminds us of the difficulties in achieving a formal nickel $Ni^{+1}$ valence state in the recently reported superconducting nickel oxides [7,8]. The history of the discovery of high-$T_c$ superconductivity would probably have been different, starting with a hypothetical $LaSrCuO_4$ compound containing $Cu^{+3}$. (We mention here that $LaSrCuO_{4-\delta}$ actually exists in our real world and is considered a useful catalyst [29], but "$\delta$" is of the order of 0.5 so that one cannot really speak of a $Cu^{+3}$ compound; the perovskite $LaCuO_3$ is metallic, but the $Cu^{+3}$ character has been questioned due to a probable Cu $3d^9$ + oxygen *p*-hole situation [30]). Regardless of whether this hypothetical fully stoichiometric starting compound $LaSrCuO_4$ is metallic or insulating, in that world one would have tried to push the valence of copper to the limits by substituting Sr by La, thereby continuously adding electrons while keeping the oxygen content constant. As the number of electrons increased and approaching the "superconducting dome" in Fig. 1 from the right side, one would eventually have arrived at a deeply metallic state for $La_{1.7}Sr_{0.3}CuO_4$. At a critical value $x_c \approx 0.25$ in $La_{2-x}Sr_xCuO_4$, the material would suddenly have become superconducting, albeit at first with a small Meissner fraction, but with the corresponding critical temperature shooting up as the number of electrons in the system increased, and without any other features accompanying this trend. What would the first conclusion have been? I am sure that most theorists had scrutinised the profoundly metallic state from which superconductivity seemed to evolve so suddenly and would have tried to identify the superconducting "glue" from the corresponding experimental data around $x_c \approx 0.25$. They might or might not have succeeded - in any case, the subsequent depression of $T_c$ beyond the maximum value with further electron addition, along with decreasing *x*, would most likely have been ascribed to undesirable effects such as charge localization and the dilution of effectively mobile charge carriers in the vicinity of a metal-insulator transition, but might at first not have been directly related to the origin of superconductivity itself.

What we may take from this thought experiment is that the terms "emerge" and "disappear" depend in some way on the prior knowledge and bias of the observer.

**c) "Underdoped" side**

I am now focusing again on the left side of the "superconducting dome" and ignore my speculation from above for now.

If the underdoped regime of cuprates is indeed governed by different effects promoting or competing superconductivity, respectively, we can search for strategies to weaken or to suppress the competing mechanisms in a controlled manner, thereby possibly increasing the critical temperature $T_c$ in this region. Applying external pressure to reduce charge ordering and at the same time to increase the critical temperature has already been demonstrated to be effective, for example, in certain chalcogenides [31]. In many perovskite oxides, which are structurally closely related to cuprates, the tendency for charge ordering can also be weakened by the application of external pressure [32,33], and it has indeed been clearly shown how pressure is affecting the charge-ordering pattern also in some cuprates [34]. It is therefore not surprising that the huge positive pressure effect on $T_c$ in cuprates [35,36] is probably partly a consequence of an associated weakening of the tendency to charge order, as stated, for example, in Refs. [35,37,38]. However, charge ordering may only be one among several other factors that determine the sign and the magnitude of the pressure effect in cuprates [36].

The transient effects observed upon irradiation of copper-oxide superconductors with ultrashort mid-infrared light pulses have sometimes also been associated with the destruction of a charge order [39]. While this would partially explain the superconductor-like properties at temperatures far above the "superconducting dome" in a natural way, similar features have also been observed in other types of superconductors such as $K_3C_{60}$ [40,41], which may hint to other possible explanations [39]. The lesser-known increases of $T_c$ and other long-lived photo-conductivity effects in cuprates [42-49] and perovskite-type manganites [50] that have been reported for a steady, continuous illumination with light, may point in a similar direction, though. The $YBa_2Cu_3O_{7-\delta}$ shows an increase in the critical temperature of up to 15 K in the underdoped region [43], which has sometimes been interpreted as a result of oxygen rearrangement at the corresponding copper-oxygen chain sites [43-46,49]. However, enhancements of the critical temperature under continuous light illumination have also been observed in other cuprates that do not contain such chains [47,48]. In any case, all of these light-induced effects might actually point to a common origin.

There may even be other ways to achieve the goal. A temporary exposure of the perovskite-type $Pr_{0.7}Ca_{0.3}MnO_3$ to 200 kV electrons in an electron microscope at $T = 92$ K, well below the respective charge-ordering temperature of $\approx 180$ K [51], has been shown to reversibly melt the ordered charge pattern. Alternatively, the persistent melting of charge order could be achieved by the exposure to X-ray radiation at sufficiently low temperature, which has also been accomplished in $Pr_{0.7}Ca_{0.3}MnO_3$ at $T = 4$ K with 8 keV radiation [52]. Such techniques, which additionally offer the possibility to study the spatial ordering simultaneously, could be combined with an *in situ* resistivity measurement as described in Ref. [52] to test a possible effect on $T_c$. Further such (admittedly challenging) experiments on underdoped cuprate superconductors would therefore be very attractive, as they could also lead to a change in the respective critical temperatures and thus give us a further taste of the intrinsic shape of the "superconducting dome" without the presence of harmful effects.

Conversely, static charge ordering could be promoted (and $T_c$ presumably reduced) by the creation of permanent lattice defects as pinning centers as discussed in Ref. [53] for NbSe$_3$. It would, however, be very difficult to disentangle the effects of the enhanced charge ordering from those coming from bare impurity scattering, which both would reduce the critical temperature to some extent.

**4. Concluding remarks**

In many synthesis experiments on different cuprate systems, optimal doping can be reached very easily, whereas heavily overdoped or completely undoped samples can often only be produced with great effort. This is especially true for those cuprates where doping is achieved by varying the oxygen content alone. The conditions for achieving an optimum transition temperature, once the cationic lattice is established, are for the most part not very sensitive to variation in subsequent heat treatments. It almost seems that a copper valence around +2.15 to +2.18 is preferred by nature, which then leads to an almost optimal transition temperature, and one may wonder whether this tendency is accidental or has some significance for the mechanism of superconductivity.

Deviations of the doping level from the optimum, downward or upward, lead to a weakening and eventual disappearance of global, phase coherent superconductivity on both sides of the dome. If competing mechanisms are partly responsible for this, there may be ways to eliminate them or at least to reduce their influence, however. At the apparent maximum of the dome in optimally doped samples, the Meissner effect is known to be largest, but outside this effect strongly decreases along with decreasing $T_c$. A Meissner effect of less than 100% (if not obscured by pinning of magnetic vortices) means that the sample is fragmented, and necessarily contains regions that cannot carry supercurrents and therefore are not superconducting. It is even conceivable that the presence and intimate mixing of these non-superconducting with the superconducting regions at the microscopic level pulls down the observable critical temperature on both sides of the dome in a manner reminiscent of the conventional or a variant of the so-called Giant Proximity Effect [54]. Beyond certain doping levels outside the superconducting dome, non-superconducting regions dominate, so that phase-coherent superconductivity is completely lost. The interpretation of inhomogeneous superconductivity, whether static or dynamic, has been proposed in different variants by several authors (see, e.g., Refs. [19,55-60]) and is probably correct.

Finally, I would like to note that for simple metallic superconductors, superconductivity does not "appear" from somewhere - the critical temperature is simply given. In this sense, an (un)conventional view to characterize the "superconducting dome" in cuprates would be that the unadulterated superconductivity "appears" exactly at optimal doping (as in the sample $x = 0.75$ reported in the seminal Ref [1] that we honor here), whatever the mechanism of electron pairing leading to superconductivity, but gradually "disappears" on both sides of the dome.